\documentclass[final,5p,times,twocolumn]{elsarticle}
\usepackage{graphicx}
\usepackage{float}
\usepackage{hyperref}
\usepackage{dsfont}
\usepackage{slashed}
\usepackage{color}
\usepackage{amsmath}
\usepackage[normalem]{ulem}
\biboptions{sort&compress}

\journal{Physics Letters B}

\def\bs{\boldsymbol}

\def\s{{\rm s}}
\def\inter{{\rm I}}

\def\med{{\rm med}}
\def\tr{{\rm Tr}}

\def\bfk{{\bs k}}
\def\bfq{{\bs q}}
\def\bfu{{\bs u}}

\def\bfp{{\bs p}}
\def\bfkp{{\bs \kappa}}

\begin{document}

\begin{frontmatter}

\title{Exploiting $\nu$-dependence of projected energy correlators in HICs}

\author[a]{Ankita Budhraja}
\affiliation[a]{organization={Nikhef Theory Group},
            addressline={Science Park 105}, 
            city={Amsterdam},
            postcode={1098 XG}, 
            country={The Netherlands}}

\author[b]{Balbeer Singh}
\affiliation[b]{organization={Department of Physics},
            addressline={University of South Dakota}, 
            city={Vermillion},
            postcode={57069}, 
            state={SD},
            country={USA}}

\begin{abstract}
We extend the recently derived factorization formula for energy-energy correlators to study the analytic structure of general $\nu$-point projected energy correlators in heavy ion collisions. The $\nu$-point projected energy correlators (or, $\nu$-correlators) are an analytically continued family of the integer $N$-point projected energy correlators, which probe correlations between $N$ final-state particles. By tracking the largest separation ($\chi$) between the $N$ particles, in vacuum, their structure is closely related to the DGLAP splitting functions and exhibits a classical scaling behavior $\sim 1/\chi$ which is modified by resummation through the anomalous dimensions. We show that, in a thermal medium, the $\nu$-correlators display non-trivial angular scaling already at the leading order in perturbation theory. We find that
for non-integer values, particularly $\nu < 1$, medium-induced jet function is enhanced compared to $\nu > 1$. 
This is particularly manifested in the ratios of $\nu$-correlators with respect to the two-point energy correlator, which encodes 
intrinsic angular information for $\nu<1$ when compared to large $\nu$ values. Moreover, for small-$\nu$ values, the $\nu$-correlators appear to saturate at $\nu=0.01$.   
We further confirm our leading-order numerical computations against simulated events from JEWEL for the parton-level production cross-section. 
Finally, we qualitatively discuss the effect of BFKL resummation for various values of $\nu$. 
\end{abstract}


\end{frontmatter}

\section{Introduction}
\label{introduction}

Jets are collimated sprays of hadrons generated through the evolution of an energetic quark/gluon (color charge parton) produced in a hard scattering event in particle collisions~\cite{Salam:2010nqg}. The evolution of this energetic parton spans through several energy scales, ranging from large perturbative to small non-perturbative, thereby encoding
the dynamics of its branching history. In heavy-ion collisions (HICs), the color charge parton in the jet undergoes further interactions with the constituents of the quark gluon plasma (QGP). These interactions manifest themselves in the form of medium-induced splittings or branchings contributing to in-medium evolution of the jet~\cite{Connors:2017ptx,Wiedemann:2000za,Blaizot:2012fh,Andres:2020vxs,Schlichting:2020lef,Casalderrey-Solana:2011ule,CMS:2017qlm}. While the initial state hard scatterings producing the jet are expected to take place at a perturbative scale, the subsequent evolution at lower energy scales gets modified through the interactions with the medium~\cite{Guo:2000nz,Mehtar-Tani:2024mvl,Blaizot:2013vha}. As a result, the final jet observed in HICs retains signatures of the various stages of the medium, see Refs.~\cite{Wiedemann:2009sh, Majumder:2010qh, Mehtar-Tani:2013pia, Blaizot:2015lma, Qin:2015srf, Apolinario:2022vzg, Budhraja:2024ttc} and references therein.

The interaction between a fast-moving color charge parton with the medium constituents occurs primarily through scattering processes, which can also trigger medium-induced radiation~\cite{Gyulassy:2000er}. One such mechanism contributing to the in-medium evolution of jets is Landau-Pomeranchuk-Migdal (LPM) effect, which is a coherent effect of multiple scattering centers in the medium~\cite{Landau:1953um,Migdal:1956tc}, see Ref.~\cite{Andres:2023jao} for recent updates. As a result, the effective transverse momentum transfer by the medium is characterized by $\hat{q}$, which is the transverse momentum squared per unit length. Since a jet consists of multiple fast-moving color charge partons, these partons can interfere among themselves, forming interference patterns which leads to color coherence dynamics characterized by a coherence angle $\theta_c$. This implies that if the two prongs of the jet are separated by an angle $\theta>\theta_c$, they can act as independent sources of medium-induced radiation. Conversely, if $\theta<\theta_c$, medium-induced radiation is coherently sourced by the total color charge of the jet. Recently, in Ref.\cite{Mehtar-Tani:2024mvl}, it was pointed out that this scale may lead to new logarithmic structures that may give rise to non-trivial evolution of the jet in the medium. Over the past two decades, numerous theoretical and phenomenological efforts have been directed at formulating an understanding of jet-medium interactions in the presence of QGP medium~\cite{Gyulassy:2003mc, Salgado:2003rv, Qin:2007rn, Zapp:2008gi, Casalderrey-Solana:2014bpa, Qin:2015srf, He:2015pra, Chien:2015hda, Wang:2016fds, He:2018xjv, Casalderrey-Solana:2019ubu, Caucal:2019uvr, Vaidya:2020cyi, Vaidya:2020lih, Baty:2021ugw, Cunqueiro:2021wls, Vaidya:2021vxu, Caucal:2021cfb, Mehtar-Tani:2021fud, JETSCAPE:2022jer, Budhraja:2023rgo, Zhang:2023oid, Barata:2023zqg, Mehtar-Tani:2024smp, Singh:2024pwr}. 

In the context of HICs, the two primary considerations for understanding medium modifications on jets are (a) what the best are observables to quantify medium-induced evolution from the vacuum evolution, and (b) whether we can develop a theoretical framework that can be systematically improved along with providing controlled theoretical uncertainties? With respect to the first question, 
energy correlators have recently emerged as promising observables to quantify medium-induced jet dynamics, as they provide a direct connection to the fundamental field-theoretic energy flow observables. Various attempts have been made to understand the structure of the simplest case of the two-point energy correlator in the medium~\cite{Andres:2022ovj, Yang:2023dwc, Andres:2023xwr, Barata:2023bhh, Andres:2024ksi, Singh:2024vwb, Barata:2024ieg, Barata:2024nqo}.
While two-point correlations provide the simplest access to the transverse scales, higher-point correlations hold the potential for understanding the rich dynamics inherent to many-body interactions in quantum chromodynamics (QCD). Recently, a few efforts have also been made to explore the structure of these higher-point energy correlators, see Refs~\cite{Bossi:2024qho, Barata:2024bmx, Barata:2025fzd}. 
Secondly, with regard to a systematic theoretical framework, 
in a series of papers in Refs.~\cite{Vaidya:2020cyi, Vaidya:2020lih, Vaidya:2021vxu, Vaidya:2021mly, Mehtar-Tani:2024smp, Singh:2024pwr}, the authors initiated the first steps toward the development of such a systematic framework together with allowing for a dynamic treatment of the medium partons. Utilizing such an effective approach, the authors of Ref.~\cite{Singh:2024vwb} established the structure of the factorization theorem, in the collinear limit, for the computation of the two-point correlator in the medium~\cite{Basham:1979gh, Basham:1978zq, Basham:1978bw, Basham:1977iq}.   

In this Letter, we further extend the factorization formula for the two-point correlator derived in Ref.~\cite{Singh:2024vwb}, to understand the general structure of $\nu$ point projected energy correlators ($\nu$-correlators) in the medium. The $\nu$-correlators can be defined as an analytic continuation of the projected (integer) $N$-point correlator~\cite{Chen:2020vvp}, which is a simple generalization of the two-point case that studies the correlations between $N$-final state particles by looking at the largest separation between them. In particular, these $N$-point projected correlators are analytically continuable in $N$, which we will denote as PE$\nu$C with $\nu > 0$ for infrared and collinear safety of the observable. This analytic continuation places all $\nu$-correlators into a single analytic family of observables. $\nu$-correlators are of particular significance due to their potential to probe the anomalous dimensions of the system. In particular, the $\nu$-correlators are formally defined as

\begin{align}
&{\rm PE}\nu{\rm C}(\chi) \equiv \frac{{\rm d}\sigma^{[\nu]}}{{\rm d} \chi} = \sum_M \int {\rm d}\sigma_X \bigg[\sum_{1\leq a_1 \leq M} {\cal W}_1^{[\nu]}(a_1)\,\delta(\chi)\, + \nonumber \\
& \hspace{8pt} \sum_{1\leq a_1 < a_2 \leq M} {\cal W}_2^{[\nu]}(a_1,a_2)\,\delta(\chi-\Delta R_{a_1,a_2}) \, + \dots + \nonumber \\
& \!\!\sum_{1\leq a_1 < .. < a_M = M}\hspace{-20pt} {\cal W}_M^{[\nu]}(a_1,..,a_M)\,\delta(\chi \!-\!{\rm max}\{\Delta R_{a_1,a_2}, .., \Delta R_{a_{M-1,M}}\}) \bigg] ,
    \label{eq:PEnuC}
\end{align}  
where $\chi$ is the projection to the largest angular separation between the particles, $M$ denotes the total number of particles in the jet,  ${\rm d}\sigma_X$ represents the production cross-section to produce the final state $X$ and $\Delta R_{ij} = \Delta\eta_{ij}^2 + \Delta\phi_{ij}^2$ are the relative distance between particles $i$ and $j$ in the rapidity-azimuth plane. The weights ${\cal W}_{1,2,\dots}$ appearing in the above equation account for correlations between subsets of particles in the jet, with ${\cal W}_1^{[\nu]}$ representing all 1-particle correlations, ${\cal W}_2^{[\nu]}$ representing all 2-particle correlations, and so on. In this Letter, we will consider the case of two parton final state and provide the explicit form of the weights ${\cal W}^{[1,2]}$ in the next section.

PE$\nu$C's provide several interesting avenues for studying the structure of QCD interactions using jets. First, in the vacuum, the distribution of PE$\nu$C's is well governed by the moments of the time-like splitting kernels. Second, the $\nu \to 0$ limit provides access to study small-$x$ physics with jets. 
For the leading-order splitting of a quark to a quark and a gluon ($q \to q + g$), the vacuum distribution in the finite $\chi$ region exhibits a simple classical scaling as $\sim 1/\chi$ which is the same for all $\nu$-correlators. In particular~\cite{Chen:2020vvp},
\begin{equation}
    {\rm PE}\nu{\rm C}(\chi) = -\frac{2^{3-\nu}\alpha_s(\mu)\, C_F}{\pi}\frac{1}{\chi}\bigg(\frac{3(\nu-1)}{(\nu+1)}-4(\gamma_E+\psi^{(0)}(\nu))\bigg),\,
\end{equation}
where $\gamma_E \approx 0.57$ is the Euler-Mascheroni constant and $\psi^{(0)}$ is the digamma function. Note that the fixed-order distribution has a divergence as $\nu\to 0$ in the term proportional to $\delta(\chi)$ which we have not shown in the above equation. See Ref.~\cite{Chen:2020vvp} for the complete form. Resummation modifies this classical scaling and introduces an anomalous scaling behavior that depends on the value of $\nu$ as ${\rm d}\sigma^{[\nu]}/{\rm d}\chi \propto 1/\chi^{1-\gamma(\nu+1)}$, where $\gamma(\nu+1)$ is the anomalous scaling dimension related to the QCD splitting functions, specifically the $\nu$-th moment of the splitting function. 

In the presence of a QGP medium where the jet parton undergoes further interactions with the thermal constituents, it is worth exploring the modification of this scaling dimension as a function of $\nu$. Similar to vacuum, one hopes to gain more insight on the structure of the anomalous dimensions and the corresponding evolution of the jet and the medium through the use of $\nu$-correlators  and our study is the first attempt in this direction.  

In this work, we restrict ourselves to the case of leading-order jet function for the medium. At this order, we take the hard function as ${\cal H} \equiv \{{\cal H}_q, {\cal H}_g\}$ as $\delta(1-x) \choose 0$, where $x$ is the energy faction of the jet initiating parton, therefore the results we obtain for the medium-induced jet function can directly be interpreted as the PE$\nu$C distributions that we are interested in.

This Letter is organized as follows: In Sec.~\ref{sec:fact}, we provide details of the factorization formula for the $\nu$-correlators in the medium. In Sec.~\ref{sec:medium-jet-func}, we present the detailed computation of the medium-induced jet function at the leading order. Here we provide a complete analytical structure by considering different phase-space regions for the transverse momentum exchanges with the medium, as well as different hierarchies. In Sec.~\ref{sec:results}, we present the numerical results and also validate the results from our analytic setup against the in-medium simulated events using JEWEL. Finally, we conclude in Sec.~\ref{sec:summary}.

\section{Factorization}
\label{sec:fact}
\subsection{Set up}
In this section, we briefly review the relevant modes and factorization formula derived in Ref.~\cite{Singh:2024vwb}. Throughout we will use light cone coordinates. To begin, we define the light-like vectors $n^{\mu}=(1,0,0,1)$ and $\bar{n}^{\mu}=(1,0,0,-1)$ which satisfy the condition $n\cdot\bar{n}=2$ and $n^2=\bar{n}^2=0$. We represent a four-vector in light-cone variables as $p^{\mu}=(p^{-},p^{+},\bfp)$ with $p^-=\bar{n}\cdot p, \, p^+=n\cdot p$ and assume that the jet moves in the -$\hat{\textbf{z}}$ direction, i.e., $p^{-} \gg \bfp \gg p^{+}$.  
To facilitate factorization, we first define the momentum scaling for the collinear jet parton as $p^{\mu}_n\sim Q(1,\lambda^2,\lambda)$ with $\lambda\sim\sqrt{\chi}$ being the power-counting parameter of the effective theory, and $Q\sim \mathcal{O}(p_T)$ is the hard scale. Next, we define momentum scaling for soft medium partons as $p^{\mu}_s\sim(Q_{\med},Q_{\med},Q_{\med})$, where $Q_{\med}$ is an intrinsic medium scale. While the exact value of $Q_{\med}$ depends on single and multiple scattering scenarios (See Ref.~\cite{Singh:2024pwr}), we restrict ourselves to the single scattering limit and assume $Q_{\med}\sim$ (2-3) GeV. The interaction between the collinear jet parton and the soft partons of the medium is mediated by the exchange of off-shell Glauber gluons $\sim Q(\lambda,\lambda^2,\lambda)$. For later convenience, we also introduce another momentum scaling of jet partons, namely collinear-soft modes $\sim Q\lambda(1,\lambda^2,\lambda)$, which contribute to the measurement. 
For more details, see Refs.~\cite{Singh:2024vwb}. 
  
\subsection{Factorization}
We begin with the total density matrix, which  includes both the medium and jet density matrices, and express its time evolution as
\begin{equation}
\rho(t)=e^{-iH t}\rho(0)e^{iHt},   
\label{eq:density}
\end{equation}
where $H=H_{0}+C(Q)l^{\mu}j_{\mu}$ is the total Hamiltonian and $H_0$ contains the collinear modes describing jet dynamics, soft modes characterizing medium physics and the jet-medium interaction terms mediated through Glauber exchanges. Moreover, $C(Q)$ is the hard operator responsible for hard interaction that creates the hard parton and $\rho(0)=|e^{+}\rangle \langle e^{-}|\otimes \rho_B$, with $\rho_B$ being thermal density matrix. For simplicity, we are working with an $e^{+}e^{-}$ initial state, with $l^{\mu}$ denoting the initial leptonic current and $j_{\mu}=\bar{\chi}_n\gamma^{\mu}\chi_n$ representing the final state quark current that initiates the jet, with $\chi_n$ being the gauge-invariant collinear quark field operator. To obtain the factorization formula for the differential distribution of $\nu$-correlators, we first impose the measurement function and expand the hard operator, $C(Q)$, in Eq.~\ref{eq:density} at leading order. This gives
\begin{align}
&{\rm PE}\nu{\rm C}(\chi)=\sum_{i_1,i_2 \dots \in {\rm jet}}\lim_{t\to \infty}\tr[\rho(t)\mathcal{M}^{[\nu]}(z_{i_1},z_{i_2},\dots)] , \nonumber\\
&=C_{\mu\delta}\lim_{t\to \infty}\int_{y_l,y_r}e^{iq\cdot(y_l-y_r)} \tr[e^{-iH_{0}t}j^{\mu}(y_l)\rho(0)\mathcal{M}^{[\nu]}j^{\delta}(y_r)e^{iH_{0}t}],
\label{eq:evolution}
\end{align}
where $q$ is the center of mass energy. Further, we have defined $C_{\mu\delta}=|C(Q)|^2L_{\mu\delta}$, with $L_{\mu\delta}$ being the leptonic tensor and the shorthand notation for integration over space-time variables as $\int_{y_l,y_r}=\int d^4y_l d^4y_r$. In the second step, the summation over the particles in the jet is implicitly implied. From here onward we will suppress $\lim_{t\to \infty}$ term and is understood to be implicit. 

In Eq.~\ref{eq:evolution}, $\mathcal{M}^{[\nu]}$ is the $\nu$-correlator measurement function, which involves projecting onto the largest separation between the partons, weighted by their energy fractions $z_i\equiv E_i/\omega$ where $\omega$ is the energy of the jet initiating parton and $E_i$ are the energies of final state particles. At the leading order, for two final-state particles $q \to q+g$, the measurement function $\mathcal{M}^{[\nu]}$ is defined as 
\begin{align}
\mathcal{M}^{[\nu]}=\sum_{a=1,2}\mathcal{W}^{[\nu]}_1(i_a)\delta(\chi)+\sum_{i_1<i_2}\mathcal{W}^{[\nu]}_2(i_1,i_2)\delta(\chi-\theta_{i_1 i_2}^2) .
\label{eq:meas}
\end{align}
Here $\theta_{i_1 i_2}$ is the angular separation between the particles and $\chi$ is the measure of the largest separation. $\mathcal{W}_1$ and $\mathcal{W}_2$ are the weight functions for the contact term ($\chi =0$) and for the case where the detectors are placed on two different particles ($\chi > 0$), respectively. Explicitly, this reads as~\cite{Chen:2020vvp}
\begin{equation}
\mathcal{W}^{[\nu]}_1(i_a)=\frac{E_{i_a}^{\nu}}{\omega^{\nu}},    
\end{equation}
\begin{equation}
\mathcal{W}^{[\nu]}_2(i_1,i_2)=\frac{(E_{i_1}+E_{i_2})^{\nu}}{\omega^{\nu}}-\sum_{a=1,2}\mathcal{W}^{[\nu]}_1(i_a),    
\end{equation}  
Now, performing an operator product expansion (OPE) to integrate out the hard modes, we rewrite Eq.~\ref{eq:evolution} in the following form
\begin{equation}
{\rm PE}\nu{\rm C}(\chi)=\frac{1}{2N_c}\sum_{i\in q,\bar{q},g}\int dx\, x^{\nu}\, {\cal H}_{i}(\omega;\mu)\, J_{i}^{[\nu]}(\omega,\chi;\mu), 
\label{eq:Pnu-med}
\end{equation}
where $\mu$ is factorization scale and we define $\omega=xQ$.
Here ${\cal H}_i$ is hard function that describes the production of the jet initiating parton and $J_i$ is the jet function that encodes the final state measurement. At leading order, since we take the hard function as $\{\delta(1-x),0\}$, the summation in Eq.~\ref{eq:Pnu-med} becomes irrelevant in our case, and the PE$\nu$C distributions are simply governed by the quark-jet functions.

At this stage, the jet function $J_i$ contains both the vacuum and medium dynamics, and is defined as 
\begin{align}
J_i^{[\nu]}&=\sum_{X}\tr\Big[\rho_B\frac{\slashed{\bar n}}{2} e^{-i H_{n\s}t}\bar {\mathcal{T}}\Big\{e^{-i\int_0^t dt' H_{G,\inter}(t')} \chi_{n,\inter}(0) \Big\}\mathcal{M}^{[\nu]}|X\rangle\nonumber\\ 
&\qquad\quad\langle X|\mathcal{T}\Big\{e^{-i\int_0^{t} dt' H_{G,\inter}(t')}\bar{\chi}_{n,\inter}(0) \Big\}e^{iH_{n\s} t}  \Big], 
\label{eq:jet}    
\end{align}
where $H_{n\s}=H_n+H_{\s}$ includes both the collinear ($H_n$) and soft ($H_{\s}$) Hamiltonian and the subscript $\inter$ implies that the operators are dressed with collinear and soft Hamiltonians. Here $H_{G,\inter}$ is the Glauber Hamiltonian that mediates the interactions between jet and medium partons and is defined as 
\begin{align}
H_G=8\pi\alpha_s\sum_{i,j\in q,g}\int d^3{\bf y}\, \mathcal{O}^{ia}_{n}({\bf y})\frac{1}{\mathcal{P}_{\perp}^2}\mathcal{O}^{ja}_{s}({\bf y}),
\label{eq:GHamiltonian}
\end{align}
where $a$ is the color index and $\mathcal{P}_{\perp}$ is the transverse momentum operator which pulls out Glauber momentum from soft operators. Furthermore, $\mathcal{O}_n$ and $\mathcal{O}_s$ are collinear and soft operators defined as
\begin{equation}
\mathcal{O}_{n}^{qa}= \bar{\chi}_{ n}t^a\frac{\slashed{\bar{n}}}{2}\chi_{n}, \hspace{1em} 
\mathcal{O}_{s}^{qa}= \bar{\chi}_{s}t^a\frac{\slashed{n}}{2}\chi_{s},    
\end{equation}
where $\chi_n$ is collinear quark operator and $\chi_s$ is soft quark operator of the medium. 
While the collinear Hamiltonian contains both collinear and collinear-soft modes~\cite{Singh:2024vwb}, soft Hamiltonian represents medium interactions. The measurement operator $\mathcal{M}^{[\nu]}$ acts on the states $|X\rangle$ and pulls out energy fractions of the partons contributing to $J_i^{[\nu]}$. Since $|X\rangle=|X_n\rangle |X_s\rangle$ involves both collinear and soft modes with $E_n\gg E_s$, the contribution of soft mode to the measurement is suppressed. 

To separate vacuum and medium-induced radiation, we further expand the Glauber Hamiltonian term in Eq.~\ref{eq:jet} and rewrite the jet function as
\begin{align}
J_{i}^{[\nu]}&=\sum_{j=0}^{N_G}J_{ij}^{[\nu]}(\omega,\chi;\mu) ,
\end{align}
where the index $j$ represents Glauber insertion index that runs from zero to $N_G$, with $j=0$ corresponding to the vacuum jet function. Contributions from all odd values of $j$ vanish, and the leading-order single scattering contribution comes from $j=2$. Moreover, when the jet and medium virtualities are widely separated, i.e., $p_n^2\gg p_s^2$, we can further re-factorize the jet function and match it to a lower medium scale. This has been discussed in more detail in Refs.~\cite{Singh:2024vwb,Mehtar-Tani:2024smp} and for multiple scattering in Ref.~\cite{Singh:2024pwr}. Since the matching function at leading order is unity, we can express the total jet function as
\begin{align}
&J_{i}^{[\nu]}=J_{i0}^{[\nu]}(\omega,\chi;\mu)+J_{i2}^{[\nu]}(\omega,\chi,L;\mu,\nu_{\rm cs}) , \nonumber\\
&=J_{i0}^{[\nu]}(\omega,\chi;\mu)+L\int \frac{d^2\bfk}{(2\pi)^2}{\bf{S}}^{[\nu]}(\omega,\chi,\bfk,L;\mu,\nu_{\rm cs})\otimes\mathcal{B}(\bfk;\mu,\nu_{\rm cs}) ,
\label{eq:diffcross0}
\end{align}
where $L$ is the length of the medium covered by the jet, $\bfk$ is Glauber momentum and $\mathcal{B}$ is the expectation value of soft operators that describe medium dynamics and is independent of final state measurement. ${\bf S}$ is the medium-induced function that captures the dynamics of jet evolution in the medium. Further, $\nu_{\rm cs}$ is the rapidity scale which appears in both ${\bf{S}}$ and $\mathcal{B}$ functions. The detailed structure of this function was obtained in Ref.~\cite{Singh:2024vwb}. To keep the expressions compact, we will drop all the renormalization scale dependence in the jet and medium functions, unless explicitly stated. For our purpose, we will use the diagrams evaluated in the Appendix of Ref.~\cite{Singh:2024vwb} and take soft limit to capture collinear-soft contributions.  Plugging back all the terms in Eq.~\ref{eq:evolution}, the differential distribution for $\nu$-correlators reads as  
\begin{equation}
{\rm PE}\nu{\rm C}(\chi)\!=\!\!\int \!\!dx\,x^{\nu}\, {\cal H}_i(\omega)\Bigg[J^{[\nu]}_{i0}(\omega,\chi)+J_{i2}^{[\nu]}(\omega,\chi,L)\Bigg]\!+\mathcal{O}(H_G^4).   
\label{eq:diffcross}
\end{equation}

The function ${\bf{S}}^{[\nu]}$ receives contribution from both the real and virtual diagrams which contribute to both the contact terms, i.e., $\delta(\chi)$ and the largest separation $\chi$.  
Therefore, ${\bf S}^{[\nu]}$ reads as 
\begin{align}
{\bf{S}}^{[\nu]}(\omega,\chi,\bfk,L)&=\sum_{i_1,i_2,\dots}\Big[{\bf{S}}_{RR}-{\bf{S}}_{VR}\Big]\mathcal{M}^{[\nu]}(z_{i_1},z_{i_2},\dots)\nonumber\\
&+\Big[{\bf{S}}_{RV}-{\bf{S}}_{VV}\Big]\delta(\chi),
\label{eq:csjet}
\end{align}
where summation runs over all the particles in the jet. Here ${\bf{S}}_{RR}$ and  ${\bf{S}}_{VR}$ correspond to real contributions from opposite and the same-side Glauber insertions, respectively. Similarly, ${\bf{S}}_{RV}-{\bf{S}}_{VV}$ are virtual diagram contributions from opposite and same side Glauber insertions. All the relevant diagrams for these contributions are evaluated in Ref.~\cite{Singh:2024vwb} (see Appendix A). 

Finally, we stress that while the factorization formula described above is derived for the $e^{+}e^{-}$ initial stage, it can be easily extended to pp and AA collisions, which would require initial-state parton distribution functions. 

\section{Medium induced jet function}
\label{sec:medium-jet-func}
We now evaluate the medium-induced jet function that governs the distribution of PE$\nu$C in a thermal medium. Additionally, we provide its analytic structure along with  $\nu$ and $\chi$ dependence, by simplifying it in certain phase-space regions. 

The contribution to the quark jet function (for $q\to q\,g$ process) from double Glauber insertions, leading to real and virtual terms (i.e., ${\bf S}_{RR}$ and ${\bf S}_{VR}$) in Eq.~\ref{eq:csjet}, takes the form~\cite{Singh:2024vwb}  
\begin{align}
({\bf{S}}_{RR}-{\bf{S}}_{VR})(\omega,\bfk,L)&=\frac{16\,{\alpha}_s}{\pi^2}\int \frac{dz}{z}\int\frac{d^2\bfq}{(2\pi)^2}\frac{\bfk\cdot\bfq}{\bfq^2(\bfq-\bfk)^2}\nonumber\\
&\Bigg[1-\frac{z\omega}{\bfkp^2 L}\sin\Big(\frac{\bfkp^2 L}{z\omega} \Big)\Bigg],    
\end{align}
where $\bfk$ is the Glauber exchange momentum with the medium constituent and $\bfq$ is transverse momentum of emitted collinear-soft gluon through interactions of the jet with the soft parton of the medium.  Moreover,   $\bfkp=\bfq-\bfk$,  ${\alpha}_s=g^2/4\pi$ and $z$ is the energy fraction carried by the gluon emitted from a  quark of energy $\omega$. Therefore, the measurement function for $\nu$-correlators defined in Eq.~\ref{eq:meas} simplifies to
\begin{equation}
\mathcal{M}^{[\nu]}=[z^{\nu}+(1-z)^{\nu}]\,\delta(\chi)+[1-z^{\nu}-(1-z)^{\nu}]\,\delta(\chi-\theta^2), 
\label{eq:measurement}
\end{equation}
where the angle $\theta=\theta_{qg}$ between the final state partons reads as 
\begin{equation}
\theta^2=\frac{\bfq^2}{[z(1-z)\omega]^2}.    
\end{equation}
Plugging these equations back in Eq.~\ref{eq:diffcross}, the medium-induced jet function for $\nu$-correlators reads as
\begin{align}
J^{[\nu]}(\omega,\chi,L)&=\frac{16\,{\alpha}_s L}{\pi^2}\int \frac{d^2\bfk}{(2\pi)^2}\int \frac{d^2\bfq}{(2\pi)^2}\int \frac{dz}{z}\frac{\bfk\cdot\bfq}{\bfq^2(\bfq-\bfk)^2}\nonumber\\
&\Bigg[1-\frac{z\omega}{\bfkp^2 L}\sin\Big(\frac{\bfkp^2 L}{z\omega} \Big) \Bigg]\Big[[z^{\nu}+(1-z)^{\nu}]\,\delta(\chi)\nonumber\\
&+[1-z^{\nu}-(1-z)^{\nu}]\,\delta(\chi-\theta^2)  \Big]\otimes \mathcal{B}(\bfk),
\label{eq:collsoft}
\end{align}
where $J^{[\nu]}$ is same as $J^{[\nu]}_{i2}$ in Eq.~\ref{eq:diffcross} with $i$ as quark. The medium function $\mathcal{B}$ which is the thermal expectation value of soft operators has been computed at leading order in Ref.~\cite{Singh:2024vwb}. In this work, for phenomenological purposes, we will use the same function. Moreover, the rapidity scale $\nu_{\rm cs}$ in Eq.~\ref{eq:diffcross0} does not explicitly appear at leading order and will require a higher-order computation of jet function, which will be attempted in the future. Even though the precise understanding of the rapidity scale requires higher order jet function, from factorization we can infer that the jet function obeys the Balitsky-Fadin-Kuraev-Lipatov (BFKL)  evolution equation and we will discuss the corresponding resummation in the next section.

For the purposes of this study, we focus on the finite $\chi$ region and hence ignore the $\delta(\chi)$ terms in Eq.~\ref{eq:collsoft} as well as suppress all the scale dependence. As mentioned earlier, besides the real diagram contributions, there are also virtual diagram contributions that do not vanish due to Glauber insertions from the medium, which only contribute to the $\delta(\chi)$ terms that we ignore for now. However, it is important to note that the delta contributions (contact terms) are crucial for obtaining the cumulative distribution.

To analyze the scaling behavior of the medium-induced jet function in a dilute medium, we start by considering two limiting cases that allow us to simplify the LPM term i.e., $\sin$ function in Eq.~\ref{eq:collsoft}. We emphasize that the approximate $\chi$-scalings we obtain are strictly valid within the single scattering approximation. For a dense medium, multiple scatterings and the incorporation of color coherence dynamics through higher-order computations of the jet function are necessary and are beyond the scope of this work. First, we consider the case when  $L\to \infty$ which reasonably describes the distributions when the system size is large. In practice, we find that this approximation works sufficiently well even for $L \sim 5$ fm in the large angular regions of the correlator,  as shown in Figure~\ref{fig:nuchiapp}.
Second, we take $\kappa^2L\ll z\omega$ which roughly corresponds to the case where the formation time, $\tau_f=z\omega/\bfq^2$, of the gluon is large i.e. $\tau_f \geq L$.  

For the first case, we simplify the jet function in a large medium extent limit 
by dropping the LPM term. Additionally, at leading order, we replace the medium function $\mathcal{B}$ with $c\,g^4 T^3/\bfk^4$ where $c$ is a constant. This is because at leading order the numerator $\mathcal{I}^q(\bfk)$ as defined in Eq.~\ref{eq:medfn} has a weak dependence on the Glauber momentum (see Ref.~\cite{Singh:2024vwb} for details). We stress that such a simple treatment of $\mathcal{B}$ may not be true if higher-order terms introduce a strong $\bfk$ dependence in $\mathcal{I}^q(\bfk)$. Nevertheless, we retain the simple leading-order structure for this study.

To begin with, we first perform the angular integration in Eq.~\ref{eq:collsoft} which evaluates to: 
\begin{align}
\int_0^{\pi} d\theta \frac{\bfk\cdot \bfq}{\bfkp^2} = \frac{\pi}{2} 
\begin{cases}
\frac{2\bfq^2}{\bfk^2-\bfq^2}, \hspace{1em} |\bfk| > |\bfq|, \\
\\[-2ex]
\frac{2\bfk^2}{\bfq^2-\bfk^2}, \hspace{1em} |\bfk| < |\bfq|.
\end{cases}
\label{eq:angint}
\end{align}
With this, we can perform the integration over the transverse momentum of the radiated parton using the measurement delta function. In the phase-space region $|\bfk|>|\bfq|$, we then obtain the following form for the jet function
\begin{align}
&J^{[\nu]}(\omega,\chi,L)\!=\!\frac{{\alpha}_s\omega^2c g^4 T^3L}{\pi^4}\!\!\!\int\!\!\!\frac{d^2\bfk}{\bfk^2(\bfk^2+m_D^2)^2}\!\! \int_{0}^{\frac{|\bfk|}{\sqrt{\chi}\omega}}\!\!\!\! dz\, (\nu z^2-z^{\nu+1})\nonumber\\
&=\!\frac{{\alpha}_sc g^4 T^3L}{3\pi^4\chi^{\frac{3}{2}}\omega}\!\!\int_{0}^{\sqrt{\chi}\omega}\!\!\!\!\frac{|\bfk|d|\bfk|}{(\bfk^2+m_D^2)^2}\bigg[|\bfk|\nu\!-\!\frac{3|\bfk|^{\nu}}{(\nu+2)(\sqrt{\chi}\omega)^{\nu-1}} \bigg].
\end{align}
Here $m_D$ is Debye mass which regulates the infrared divergences in the medium. The appearance of the Debye mass in the Glauber propagator can be understood as soft loop contributions in it. Note that in Eq.~\ref{eq:jetfnexp}, to be consistent with the collinear-soft modes in the effective theory description, we have also expanded the weight function in the soft limit.

We note that in this case the jet function is enhanced by the length of the medium in contrast to the second case, which we discuss shortly. Moreover, the upper limit on $|\bfk|$ integration arises from the constraint on the momentum fraction $z<1$. Upon integrating over the Glauber momentum, the medium-induced jet function acquires the form    
\begin{align}
&J^{[\nu]}(\omega,\chi,L)=\frac{{\alpha}_sc g^4 T^3L}{3\pi^4\chi^{\frac{3}{2}}\omega}\bigg[\bigg(\frac{1}{m_D}\tan^{-1}\bigg(\frac{\sqrt{\chi}\omega}{m_D} \bigg)-\frac{\sqrt{\chi}\omega}{(\chi\omega^2 + m_D^2)}\bigg) - \nonumber \\
&\frac{3\chi^{\frac{3}{2}}\omega^3}{2(\nu+2)m_D^{2}} \bigg\{\frac{1}{(\chi\omega^2 + m_D^2)}-\!\frac{\nu \, {}_2F_1(1,1+\frac{\nu}{2};2+\frac{\nu}{2};-\frac{\chi\, \omega^2}{m_D^2})}{m_D^2(\nu+2)}\bigg\}\bigg],  
\label{eq:jetmed1}
\end{align}
where ${}_2F_1$ is the hypergeometric function. For realistic values of $\omega$, $\chi$ and $m_D$ considered here, we expand the expression above in terms of $m_D^2/(\chi\omega^2)$, which yields a leading scaling behavior of the medium-induced jet function as
\begin{align}
&J^{[\nu]}(\omega,\chi,L)
\approx \frac{{\alpha}_s g^4 T^3(5+4\nu)}{(\nu+2)\pi^4} \frac{L}{\chi^2\omega^2} + {\cal O}\Big(\frac{m_D^2}{\chi\omega^2}\Big),
\end{align}
where $m_D$ dependent terms are contained in the sub-leading contributions. We observe that, in this phase-space regime, the leading behavior of the medium jet function behaves as $1/\omega^2$ with the jet energy, and as $1/\chi^2$ with respect to the measurement variable. 
It is important to emphasize that such a leading structure only appears as a result of the expansion and the complete analytic structure of the medium-induced jet function is much more non-trivial in both $\chi$ and $\nu$ as shown in Eq.~\ref{eq:jetmed1}.

Next, we consider the second phase-space region in Eq.~\ref{eq:angint} and simplify the jet function given in Eq.~\ref{eq:collsoft} with $|\bfk|<|\bfq|$. After performing the $\bfq$ integration similar to the previous case we obtain
\begin{align}
&J^{[\nu]}(\omega,\chi,L)=\!\frac{{\alpha}_scg^4 T^3 L}{\pi^4 \omega^2\chi^2}\!\!\int_{0}^{\omega\sqrt{\chi}} \!\!\! \frac{d|\bfk|\bfk^3}{(\bfk^2+m_D^2)^2} \!\int_{\frac{|\bfk|}{\omega\sqrt{\chi}}}^1 \! dz \,\bigg[\frac{\nu}{z^2}-z^{\nu-3}\bigg]\nonumber\\
&\quad=\frac{{\alpha}_scg^4 T^3 L}{\pi^4 \omega^2\chi^2} \int_{0}^{\omega\sqrt{\chi}}  \frac{d|\bfk|\bfk^3}{(\bfk^2+m_D^2)^2} \bigg[\frac{2\nu-\nu^2-1}{\nu-2} +\frac{\nu\omega\sqrt{\chi}}{|\bfk|} \nonumber\\
&\qquad\qquad\, +\frac{\chi^{1-\nu/2}|\bfk|^{\nu-2}}{(\nu-2)\omega^{\nu-2}}\bigg],
\label{eq:angint}
\end{align}
where the lower limit of the $z$ integration is set by the condition $|\bfk|<z\sqrt{\chi}\omega$ imposed by angular integration in Eq.~\ref{eq:angint}.  Finally, after integrating over Glauber momentum, we obtain
\begin{align}
&J^{[\nu]}(\omega,\chi,L)=\frac{{\alpha}_scg^4 T^3 L}{2\pi^4 \omega^2\chi^2} \bigg[\frac{2\nu-\nu^2-1}{\nu-2}\bigg(\log\bigg[1+\frac{\chi\omega^2}{m_D^2}\bigg]  \nonumber\\
&\quad -\frac{\chi\omega^2}{(\omega^2\chi+m_D^2)}\bigg)+ \nu\omega\sqrt{\chi} \bigg(\frac{1}{m_D}\tan^{-1}\bigg(\frac{\sqrt{\chi}\omega}{m_D}\bigg) \nonumber\\
&\quad -\frac{\sqrt{\chi}\omega}{(\omega^2\chi+m_D^2)}\bigg) +\frac{\chi^2\omega^4}{(\nu-2)m_D^4}\bigg(\frac{m_D^2}{(m_D^2+\chi\omega^2)} \nonumber \\
&\quad -\frac{\nu \, {}_2F_1(1,1+\frac{\nu}{2};2+\frac{\nu}{2};-\frac{\chi\, \omega^2}{m_D^2})}{(\nu+2)}\bigg)\bigg] .
\end{align}
First, we note that there is no divergence at $\nu \to 2$ as might be naively expected from the above equation. This is because for $\nu \to 2$, the hypergeometric function reduces to a special form that can be expressed in terms of a logarithmic function, which combines with the first term to yield a finite result. Similar to the previous case, we expand this equation in terms of the parameter $m_D^2/(\chi\omega^2)$ to obtain
\begin{align}
&J^{[\nu]}(\omega,\chi,L)
\approx \frac{{\alpha}_sg^4 T^3 \nu L}{\pi^4 \omega\, m_D \chi^{3/2}} \, .
\end{align}
where we have dropped subleading terms along with terms of $\mathcal{O}(m_D^2/(\chi\omega^2))$. Once again, we observe that the medium-induced jet function is suppressed with the jet parton energy as $1/\omega$ as well as $\chi$. However, in this case, there is also an overall enhancement due to the Debye mass $m_D$. 
As a result, the contribution to the total jet function in the limit $L \to \infty$ is dominated by the phase-space region where $|\bfk|<|\bfq|$. Note that there is also an interplay between the two phase-space regimes as $\nu$ approaches zero. The leading expansion suggests that as $\nu \to 0$, the dominant contribution arises from the region where $|\bfk|>|\bfq|$. 

\begin{figure}[t]
\centering 
\includegraphics[width=0.43\textwidth]{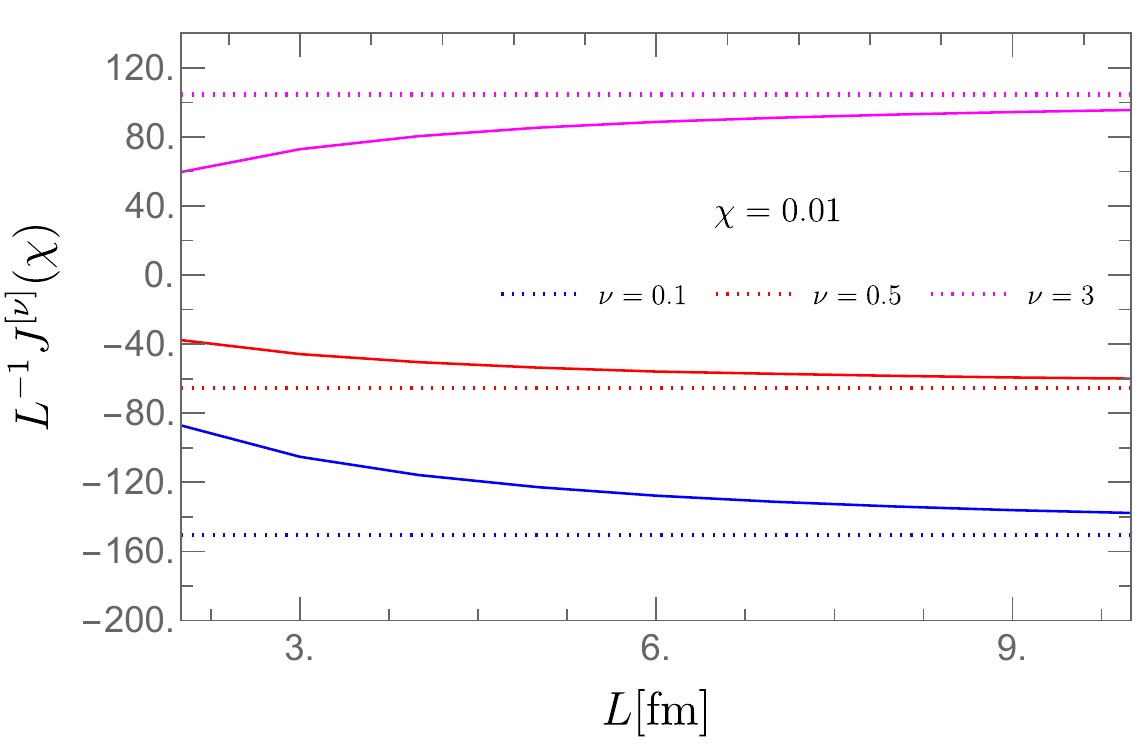}	
\caption{Variation of the jet function as a function of $L$ for various values of $\nu$ at fixed $\chi = 0.01$. The solid lines show the results obtained from the full numerical computation of the jet function and dotted lines represent the approximation $L \to \infty$. The medium parameters are $T=0.4$ GeV, $L=5$ fm and $m_D=0.8$ GeV. } 
\label{fig:nuchiL}
\end{figure}

In Figure~\ref{fig:nuchiL}, we compare the approximate results in the large system size limit $L \to \infty$ against the full numerical results obtained by keeping the entire phase space, as shown later in Section~\ref{sec:results}. We present this as a function of $L$ for different values of $\nu$, namely $\nu = 0.1,0.5$ and $3$ and for a fixed value of $\chi = 0.01$. The solid curves show the full numerical results, and the dotted curves represent the $L \to \infty$ approximation. The jet function is normalized by the overall factor of the length of the medium, appearing in Eq.~\ref{eq:diffcross} and the constant $c$ is taken as ${\cal O}(1)$ in both cases. As evident from the figure, the approximate results and the scaling behavior discussed above hold well when the system size is sufficiently large, i.e. $L \gtrapprox 10$ fm for $\chi \sim 0.01$. The results as a function of $\chi$ are further shown in ~\ref{apn:plotexp}.

Next, for the case where the gluon formation time is large, we expand the $\sin$ function in the limit where $\kappa^2L\ll z\omega$. We emphasize that this condition only approximately corresponds to the formation time and is not an exact equivalence. In the numerical results that we provide in Sec.~\ref{sec:results}, we do not perform any expansions and evaluate the full integration numerically. Nevertheless, we present a comparison of the approximate computations we perform in this section against our full numerical estimations in ~\ref{apn:plotexp}. Upon expanding the $\sin$ term, the jet function reads as
\begin{align}
&J^{[\nu]}(\omega,\chi,L)=\frac{8{\alpha}_sL^3}{3\pi^2}\int\frac{d^2\bfk}{(2\pi)^2}\int\frac{d^2\bfq}{(2\pi)^2} \int \frac{dz}{z} \frac{(\bfk\cdot\bfq)\bfkp^2}{\bfq^2}\nonumber\\
&\quad\, (\nu z-z^{\nu})\delta(\bfq^2-\chi z^2\omega^2)\Theta(z\omega-\bfkp^2 L)\otimes\mathcal{B}(\bfk),
\end{align}
where we have dropped terms of $\mathcal{O}(L^4)$ and higher. 
With this simplification, we perform the integration over the transverse momentum of the radiated parton using the measurement delta function to acquire
\begin{align}
&J^{[\nu]}(\omega,\chi,L)= \frac{4c{\alpha}_sL^3 g^4T^3}{3\pi^2}\int \frac{d^2\bfk}{(2\pi)^2}\frac{1}{(\bfk^2+m_D^2)^2}\nonumber\\
&\int dz\int\frac{d\theta}{(2\pi)^2}\frac{|\bfk|\bar{\bfq}\cos\theta\,\bar{\bfkp}^2}{\bar{\bfq}^2}\Big\{\nu-z^{\nu-1}\Big\}\Theta(z\omega-\bar{\kappa}^2L), 
\label{eq:jetfnexp}
\end{align}
where $\bar{\bfq}=z\omega\sqrt{\chi},\, \bar{\bfkp}^2=\bfk^2+\bar{\bfq}^2-2|\bfk|\bar{\bfq}\cos\theta$.
Next, to simplify the integration over the solid angle we impose the constraint   
$|\bfk|\ll 1/(2\sqrt{\chi}L)$ from the $\Theta$ function to obtain 
\begin{align}
J^{[\nu]}(\omega,\chi,L)= \frac{c{\alpha}_sL^3 g^4T^3}{6\pi^4}\!\int_{0}^{\frac{1}{2\sqrt{\chi}L}}\!\!\frac{\bfk^3d|\bfk|}{(\bfk^2+m_D^2)^2}\int_{z_2}^{z_1}\!\! dz \{\nu-z^{\nu-1} \} ,  
\label{eq:jetz}
\end{align}
where 
\begin{align}
&z_1=\frac{1}{2}\bigg(\frac{1}{\chi\omega L}+\sqrt{\frac{1}{\chi^2\omega^2 L^2}-\frac{4\bfk^2}{\chi\omega^2}}    \bigg)\equiv\frac{1}{\chi\omega L}-\frac{\bfk^2 L}{\omega}, \nonumber\\
&z_2=\frac{1}{2}\bigg(\frac{1}{\chi\omega L}-\sqrt{\frac{1}{\chi^2\omega^2 L^2}-\frac{4\bfk^2}{\chi\omega^2}}    \bigg)\equiv \frac{\bfk^2 L}{\omega}.
\end{align}
Here we keep only the dominant term for simplicity and drop higher order terms suppressed by $\mathcal{O}(\chi \bfk^2 L^2)^2$ in both $z_1$ and $z_2$.  
After performing all the remaining integrations, the medium-induced jet function reads as
\begin{align}
&J^{[\nu]}(\omega,\chi,L)= \frac{c{\alpha}_sL^3 g^4T^3}{12\pi^4}\Bigg[\frac{2\nu L}{\omega}\bigg(\frac{1+8\chi r^2}{8\chi L^2+32\chi L^2 r^2}-m_D^2\nonumber\\
&\times\log\bigg(1+\frac{1}{4\chi r^2} \bigg) \bigg)-\frac{\nu}{2\chi\omega L}\bigg(\log\bigg(1+\frac{1}{4\chi r^2} \bigg)-\frac{1}{1+4\chi r^2} \bigg)\nonumber\\
&\qquad-\frac{L^{\nu}}{\nu \omega^{\nu}}\bigg(1-\frac{1}{1+4\chi r^2}-(\nu+1)\Gamma(\nu+2)\nonumber\\
&\,\times{}_2\tilde{F}_1(1,\nu+2;\nu+3;-\frac{1}{4\chi r^2})\bigg)+\frac{1}{\nu (\chi\omega L)^{\nu}}f(\nu,\chi,L)
\Bigg],  
\label{eq:hie2}
\end{align}
where ${}_2\tilde{F}_1$ is the regularized hypergeometric function and $\Gamma(\nu)$ is the gamma function. The function $f(\nu,\chi,L)$ can be obtained after numerical integration of
\begin{equation}
f(\nu,\chi,L)=\int_0^{\frac{1}{2\sqrt{\chi}L}} d|\bfk| \frac{|\bfk|^3 (1-\chi \bfk^2 L^2)^{\nu}}{(\bfk^2+m_D^2)^2}.    
\end{equation}
In Eq.~\ref{eq:hie2} to keep expressions compact we have defined the dimensionless variable $r=L m_D$. We note that in the limit $\nu\to 1$ the above result vanishes, as anticipated from Eq.~\ref{eq:measurement}. 
Further, as $\nu\to 0$ the $1/\nu$ terms appearing in Eq.~\ref{eq:hie2} exactly cancel each other leaving a finite term that reads  
\begin{align}
&J^{[\nu=0]}(\omega,\chi,L)=-\frac{c{\alpha}_sL^3 g^4T^3}{24\pi^4}\log(\chi\omega L)\bigg(\log\bigg(1+\frac{1}{4\chi r^2} \bigg)\nonumber\\
&\qquad\qquad\quad -\,\frac{1}{1+4\chi r^2}\bigg) , \nonumber \\
&\qquad\equiv -\frac{c{\alpha}_s g^4T^3}{768\pi^4m_D^4 L\,\chi^2}\log(\chi\omega L)+\mathcal{O}(1/\chi L^2m_D^2)^3,
\label{eq:jethie2}
\end{align}
where in the second line we have expanded the full expression in the limit $\chi L^2m_D^2\gg 1$. It is worth emphasizing that, even from the simple limits shown here, we observe that the medium-induced jet contributions saturate as $\nu \to 0$. 

In the next section, we provide our full result by solving Eq.~\ref{eq:collsoft} numerically for a static medium of length $L$ and by taking the leading order result for the medium function $\mathcal{B}$. Finally, we also present results for events simulated from JEWEL. 

\section{Results}
\label{sec:results}
For all the plots except for Figure~\ref{fig:nuchijewel}, we take medium temperature to be $T=0.4$  GeV and medium length $L=5$ fm. 
For Figure~\ref{fig:nuchijewel}, we set the medium parameters such that the jet $R_{\rm AA} \sim 0.45-0.5$ for a jet with $p_T \sim {\cal O}(100\, {\rm GeV})$. 

The leading order medium function $\mathcal{B}$ was computed in Ref.~\cite{Vaidya:2020cyi,Singh:2024vwb}, which for thermal quarks takes the form
\begin{align}
\mathcal{B}(\bfk)=\frac{(8\pi\alpha_s)^2 }{\bfk^4} \mathcal{I}^q(\bfk),   
\label{eq:medfn}
\end{align}
where $\alpha_s$ is the coupling between the Glauber gluon and soft partons in the medium. The function $\mathcal{I}^q(\bfk)$ reads as
\begin{align}
\mathcal{I}^q(\bfk)&=\int\frac{dq^-d\bfq}{(2\pi)^3}\frac{\bfq^2}{(q^-)^2}\tilde{f}\Big(\frac{q^-}{2}+\frac{\bfq^2}{2q^{-}} \Big)\bigg[1-\tilde{f}\Big(\lambda+\frac{\bfq^2}{2q^-} \Big)\bigg] ,  
\end{align}
where $\tilde{f}$ is Fermi-Dirac distribution function and 
\begin{equation}
\lambda=\frac{q^-(\bfk+\bfq)^2}{2\bfq^2}.    
\end{equation}
For numerical purposes, we will take both quark and gluon contributions in the medium function evaluated in Ref.~\cite{Singh:2024vwb}.

\subsection{Numerical analysis}
\label{res:numerics}
\begin{figure}[h]
\centering 
\includegraphics[width=0.43\textwidth]{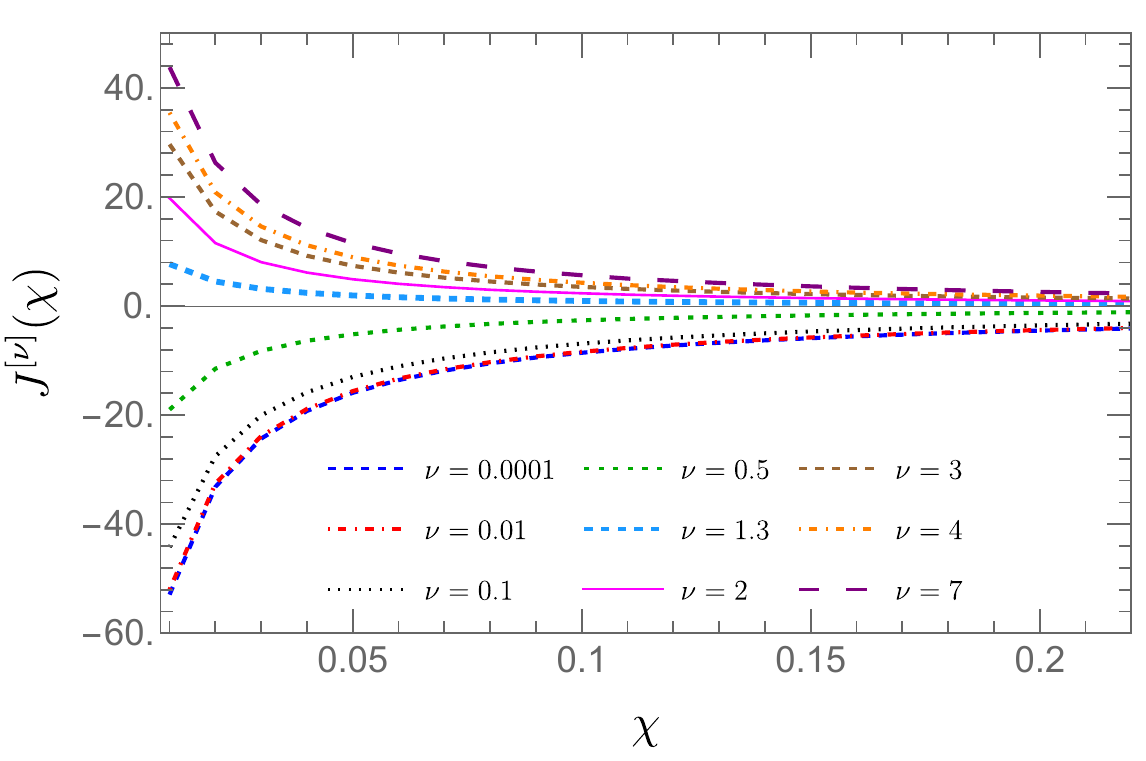}	
\caption{Variation of the jet function as a function of angle $\chi$ for various values of $\nu$. The medium parameters are $T=0.4$ GeV, $L=5$ fm and $m_D=0.8$ GeV. } 
\label{fig:nuchi}
\end{figure}

With this setup for the medium function and the medium parameters, we first show the variation of the jet function with angular separation between the final state particles, $\chi$, for various values of $\nu$ in Figure~\ref{fig:nuchi}. We take the medium coupling to be $g=2$. We note that for $\nu<1$, the fixed-order jet function saturates near $\nu=0.01$ whereas in vacuum there is an explicit $1/\nu$ enhancement~\cite{Chen:2020vvp}.
Furthermore, for $\nu<1$ the jet function slowly approaches zero compared to $\nu>1$. As a consequence of this, the fixed order ratios of $\nu$-correlators with standard two point energy correlator are enhanced for $\nu<1$, which we will discuss shortly.  
\begin{figure}[h]
\centering 
\includegraphics[width=0.43\textwidth]{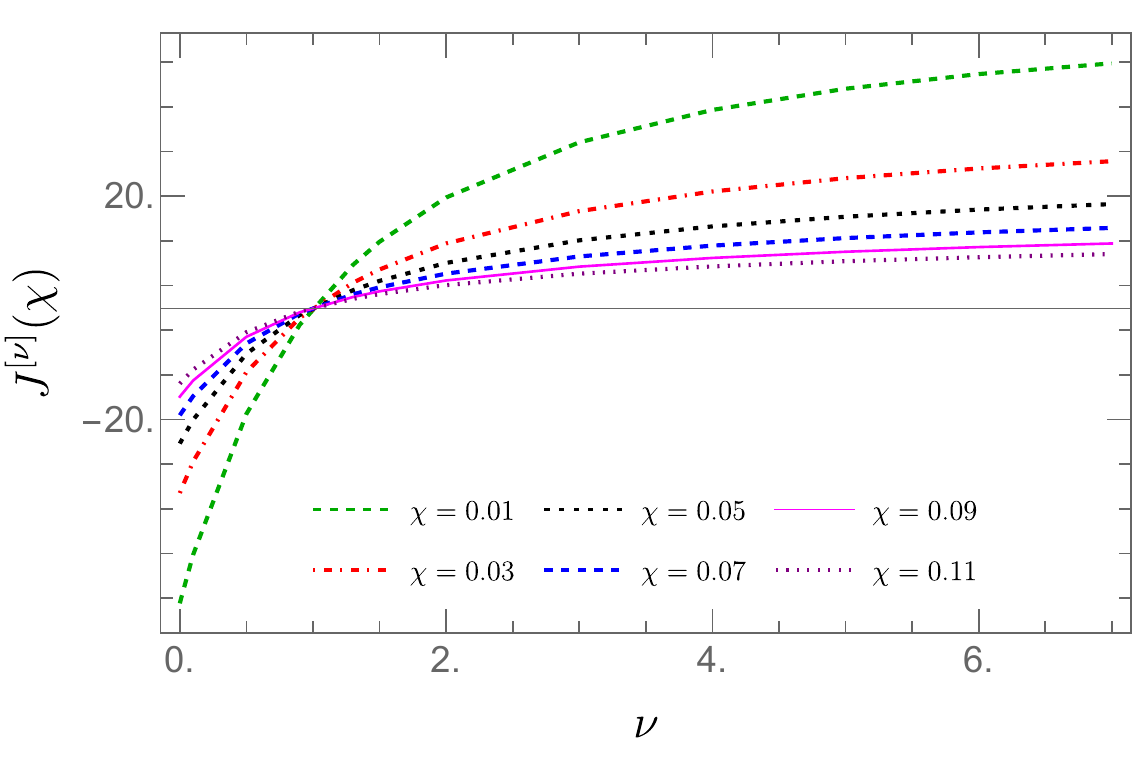}	
\caption{Variation of jet function as a function $\nu$ for various values of $\chi$. Here again the medium parameters are $L=5$ fm, $T=0.4$ GeV. } 
\label{fig:nuplot}
\end{figure}

Next, in Figure~\ref{fig:nuplot}, we plot the variation of the medium-induced jet function as a function of $\nu$ for various values of $\chi$ ranging from small to large. We observe that, similar to the vacuum case, the PE$\nu$C distributions become negative for the medium-induced case when $\nu<1$.\footnote{Note that the cumulative distributions are still positive due to the large positive contribution associated to the contact contribution and we expect the same to hold for the case of medium-induced distributions as well.} However, unlike the vacuum case, this cannot be trivially mapped to the anomalous dimensions and we stress that such a mapping demands higher-order computations where we can explicitly see anomalous dimensions  in the medium  

\begin{figure}[h]
\centering 
\includegraphics[width=0.48\textwidth]{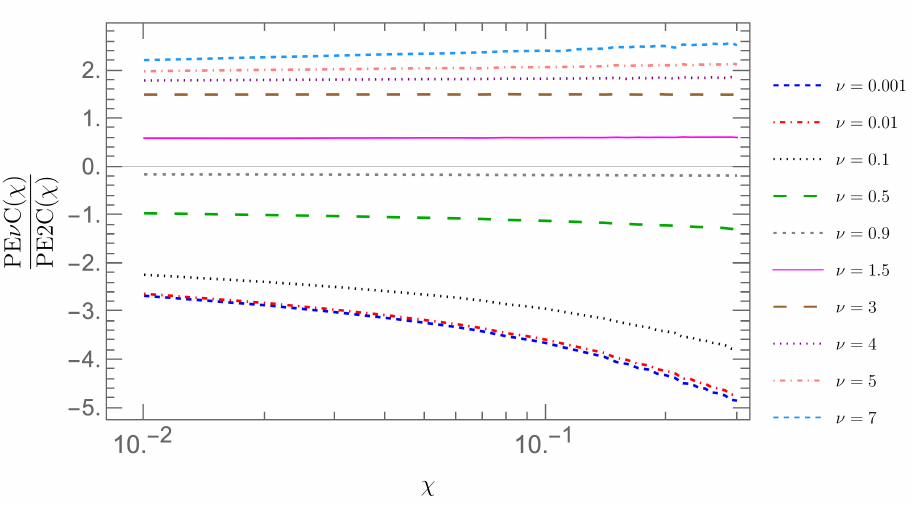}	
\caption{Variation of ratios of $\nu$-point energy correlators with that of $\nu=2$ as a function of angle $\chi$. Medium parameters are same as the one used in Figure~\ref{fig:nuratio}.  } 
\label{fig:nuratio}
\end{figure}

Finally, to assess the relevance of $\nu$-correlators for various integer and non-integer values of $\nu$ in addition to the two-point correlators, we plot the ratios PE$\nu$C/PE2C at leading order for various values of $\nu$ in Figure~\ref{fig:nuratio}. We find that for all values of $\nu>1$, the higher-point correlators are enhanced in the medium across the entire range of angular scales. Additionally, we observe a mild angular dependence in the ratio of these correlators when compared to the two-point case.
In contrast, correlators with $\nu<1$ are much more interesting in the medium. For these small-$\nu$ values, we observe that not only the medium effects are enhanced but an intrinsic non-trivial dependence on the angular scale of the correlator also arises. In Sec.~\ref{sec:JEWEL}, we also validate this finding against Monte Carlo results obtained from JEWEL. Next, we analyze the impact of resummation on the results of the fixed order distributions presented so far.   

\begin{figure}[t]
\centering 
\includegraphics[width=0.43\textwidth]{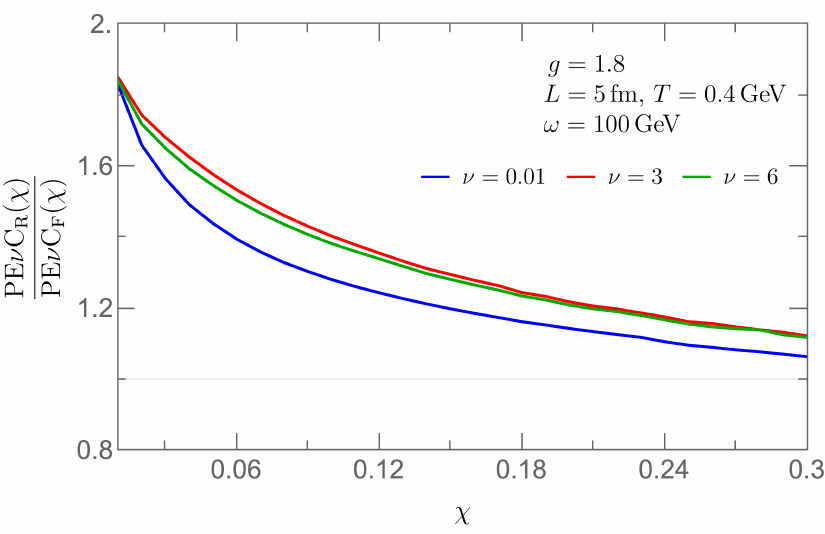}	
\caption{Variation of ratios of resummed and leading order jet functions as a function of angle $\chi$ for various values of $\nu$ and $\omega=100$ GeV.} 
\label{fig:resum}
\end{figure}

 From renormalization group (RG) consistency, we can infer that medium-induced jet function obeys a BFKL evolution equation with anomalous dimension same as the medium function but with an opposite sign~\cite{Vaidya:2021vxu}. 
However, higher order computations of the jet function are necessary for the knowledge of the precise scale relevant for BFKL evolution. To obtain the renormalized jet function we follow the prescription outlined in Ref.~\cite{Kovchegov:2012mbw} to solve
\begin{align}
\frac{d{\bf S}^{[\nu]}(\bfk,\nu_{\rm cs})}{d\ln\nu_{\rm cs}}\!=-\frac{\alpha_s(\mu) N_c}{\pi^2}\!\!\!\int \!d^2\bfu\bigg[\frac{{\bf S}^{[\nu]}(\bfu,\nu_{\rm cs})}{(\bfu-\bfk)^2}-\frac{\bfk^2{\bf S}^{[\nu]}(\bfk,\nu_{\rm cs})}{2\bfu^2(\bfu-\bfk)^2} \bigg],
\label{eq:resumjet}
\end{align}
where as mentioned earlier $\nu_{\rm cs}$ is the scale associated with rapidity renormalization and $\mu$ is the UV scale respectively. In order to solve the above evolution equation, one needs the precise knowledge of the rapidity scale for the jet function which for our case only appears at next-to-next-to-leading order (NNLO) which has not been computed so far. Nevertheless, to qualitatively capture the impact of resummation, we perform running from a scale $\nu^{0}_{\rm cs} \equiv Q_{\med}/\sqrt{\chi}$ to the medium scale $Q_{\med}\sim 3$ GeV. This yields % and obtain
\begin{align}
{\bf S}^{[\nu]}_{\rm R}(\bfk)=\frac{1}{\pi |\bfk|}\sqrt{\frac{\pi}{14\zeta(3)\bar{\alpha}Y}}e^{(a_p-1)Y}\int d^2\bfu \frac{{\bf S}^{[\nu]}(\bfu)}{|\bfu|} e^{-\frac{\log^2(|\bfk|/|\bfu|)}{14\zeta(3)\bar{\alpha}Y}} ,
\label{eq:jetresum}
\end{align}
where $a_p=1+4 \bar{\alpha} \ln{2}$, $Y=\log(\nu^{0}_{\rm cs}/|\bfk|)$ and $\bar{\alpha}=\alpha_s N_c/\pi$.
The transverse momentum scale $|\bfk| \sim Q_{\med}$ so that $Y \sim \ln 1/\sqrt{\chi}$. We plot the variation of the ratio of resummed and leading jet function with $\chi$ in Figure~\ref{fig:resum}. From the plot, we note a very mild dependence of the resummed $\nu$-correlators on $\nu$. This is because, for resummed jet function, the only $\nu$ information is in the boundary condition (in Eq.~\ref{eq:resumjet}) which is the fixed order jet function. However, in more realistic scenarios we also need the information of $\nu$ dependence of anomalous dimensions which only appears at NNLO.   

\subsection{JEWEL Analysis}
\label{sec:JEWEL}

Having established the features of the $\nu$-point projected correlators in our analytic framework, we now validate our results against Monte Carlo simulations generated with JEWEL~\cite{Zapp:2008gi, Zapp:2012ak, Zapp:2013vla, KunnawalkamElayavalli:2017hxo} in an idealized setup (without recoils). The realistic case incorporating the default medium model in JEWEL as well as medium recoils is discussed in ~\ref{apn:fulljewel}. 
In order to establish a fair comparison, we switch off all initial-state radiation, final-state radiation and multiple-parton interaction effects within PYTHIA 6.4 complemented with JEWEL. Further, we utilize a brick setup of the medium and turn off the hadronization effects. The $\nu$-correlators are then computed on simulated events for $\sqrt{s_{NN}} = 5.02$ TeV Pb+Pb collisions with a brick of length $L=9 \, {\rm fm}$ and temperature $T = 0.5\, $ GeV. Finally,  we construct anti-kt jets with radius $R=0.4$ within the transverse momentum range $p_T \in [100,120]\, {\rm GeV}$ and psuedorapidity $\vert \eta \vert < 1.9$. For practical purposes, we use the new method detailed in Ref.~\cite{Alipour-fard:2024szj} and available through Refs.~\cite{FASTEEC, github:RENC}, for the computation of $\nu$-correlators for various values of $\nu$.

\begin{figure}[h]
\centering 
\includegraphics[width=0.47\textwidth]{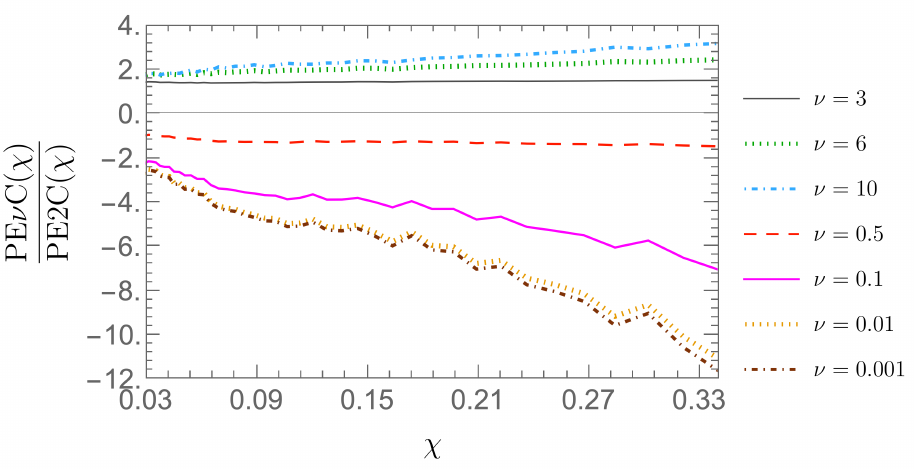}	
\caption{Ratios of PE$\nu$C distributions with respect to the 2-point distribution utilizing the data generated from JEWEL simulations.  } 
\label{fig:nuchijewel}
\end{figure}

The results for the ratios of PE$\nu$C to the two-point energy correlator are presented in Figure~\ref{fig:nuchijewel}. From the figure, we note very similar qualitative features as observed in our analytic computation, namely the limit $\nu \to 0$ exhibits an intrinsic angular dependence beyond what is captured by the two-point energy correlator.

\section{Conclusions}
\label{sec:summary}

In this Letter, we compute the medium modifications on the general structure of the $\nu$-correlators in the EFT framework within the leading order double Glauber insertions approximation. We show that $\nu$-correlators with small $\nu$ values, i.e., $\nu<1$ are enhanced, particularly in the large angle region, compared to two point energy correlators. Moreover, this enhancement is more prominent than any other higher point projected energy correlators. We further confirm this observation using simulated events from JEWEL which also accounts for multiple scatterings. For partonic level, the ratios of $\nu$-correlator against standard two point energy correlator is shown in Fig.~\ref{fig:nuchijewel}. Further, including hadronization effects as well as medium recoil and longitudinal expansion, the same distributions are plotted in Fig.~\ref{fig:nujewelfull}. 

In a dilute medium within single scattering approximation, we show the analytic structure of scaling behavior of these correlators in two limiting regimes. We first consider large spatial extent of the medium and take $L\to \infty$ limit which reasonably describes large angle region as shown in Fig.~\ref{fig:nuchiL} and Fig.~\ref{fig:nuchiapp}. Although $\nu$-correlators exhibit a classical scaling that goes like $1/\chi$ for all $\nu$-values in the vacuum, their structure is significantly more complex in the case of the medium modified jets even in single scattering limit. Additionally, in the vacuum, $\nu$-correlators are also interesting from a phenomenological point of view as they not only provide access to the leading twist collinear dynamics in jets but also allow the study of small-$x$ dynamics using jets. In HICs, we show that these correlators are interesting as they contain more angular information and hold the potential to unravel the medium effects through the use of the full $\nu$ dependent structure of these correlators.

We stress that while these small-$\nu$ values are of interest, the complete potential of these correlators requires more detailed theoretical and phenomenological efforts and we hope our study provides a motivation towards such future investigations. In particular, for realistic scenarios, correlators with $\nu<1$ will also have significant non-perturbative contributions as shown in Ref~\cite{Budhraja:2024tev} for pp collision. A similar study for HICs will be necessary to quantify the size of non-perturbative contributions in the final state distribution. Moreover, for a dense medium multiple scatterings are important and the corresponding analysis requires incorporation of multiple scattering within the EFT framework. We leave these detailed theoretical and phenomenological investigations for future study.

It is worth emphasizing that in the vacuum, the scaling of PE$\nu$C distributions is governed by the anomalous dimensions of the theory. However, in HICs, the scaling observed in the ratios of Figure~\ref{fig:nuratio} is not (yet) connected to medium evolution through anomalous dimensions but is inherently due to the non-trivial structure of the leading order medium-induced jet function. Higher-order computations of the medium jet function are therefore necessary to truly exploit the structure of the medium evolution.
Finally, we stress that, in this work, we restrict to the leading order jet function within the double Glauber insertion limit, a more detailed study that incorporates higher-order contributions will certainly improve our understanding of jet dynamics in HICs. 

\section*{Acknowledgements}
We thank Varun Vaidya, Yacine-Mehtar Tani and Felix Ringer for valuable discussions. B.S is supported by startup funds from the University of South Dakota and by the
U.S. Department of Energy, EPSCoR program under contract No. DE-SC0025545. A.B. is supported by the project `Microscopy of the Quark
Gluon Plasma using high-energy probes’ (project number VI.C.182.054) which is partly financed by the Dutch Research Council (NWO).

\appendix

\section{$\nu$-point angular Scalings}
\label{app:fit}
\begin{figure}[t]
\centering 
\includegraphics[width=0.41\textwidth]{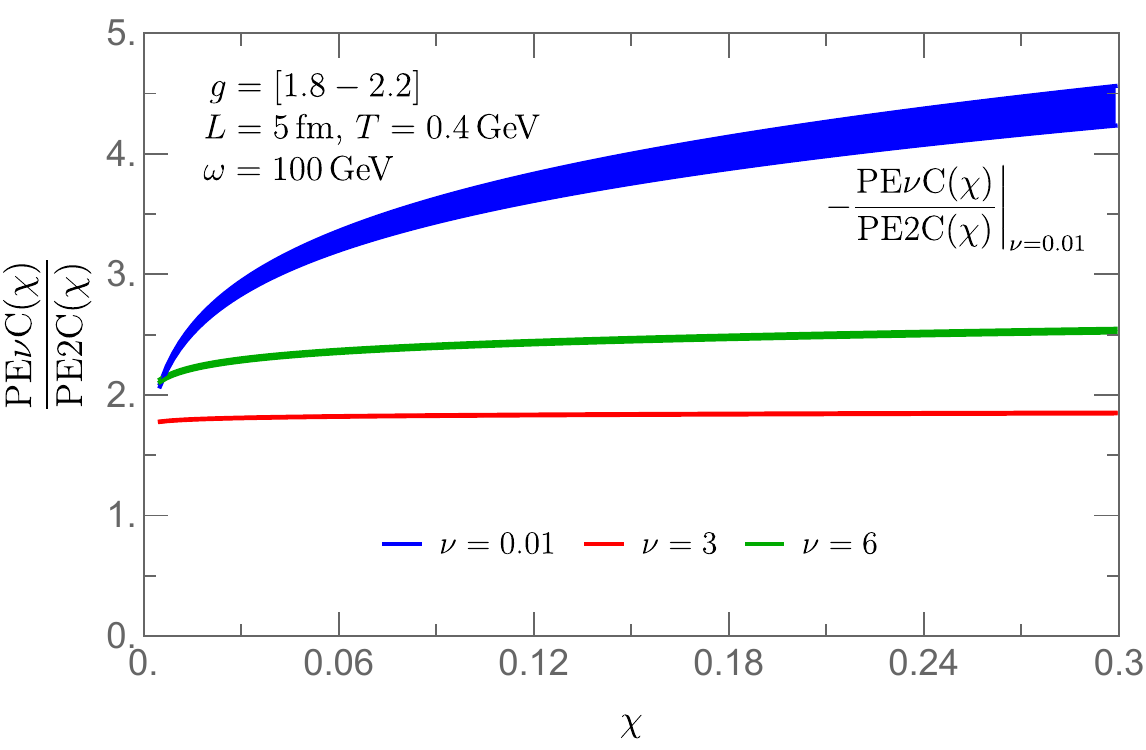}	
\caption{Approximate angular scaling behavior of $\nu$-point energy correlators for $\nu=0.01,3,6$ compared to $\nu=2$.  Medium parameters are same as the one used in Figure~\ref{fig:nuplot}.  } 
\label{fig:nuratiofit}
\end{figure}

To obtain a simple angular scaling behavior of the ratios of PE$\nu$C distributions shown in Figure~\ref{fig:nuratio}, we fit these ratios to a power law of the form $a\chi^b$. The fit values for various $\nu$ values are tabulated in Table~\ref{tab:fit}. 
\begin{table}[h]
\centering 
\begin{tabular}{ |c|c| c| c| }
\hline
 $\nu$ & a & b & $\sim\chi^{b}$ \\ 
 \hline
 0.01 & -5.74 & 0.19 & $\chi^{0.19}$ \\  
 \hline
 3 & 1.87 & 0.01 & $\chi^{0.01}$ \\ 
 \hline
 6 & 2.68 & 0.04 & $\chi^{0.04}$ \\
 \hline
\end{tabular}
\label{tab:fit}
\caption{Approximate scaling for ratio of $\nu$-correlators with that of 2-point energy correlators. Here we take the same medium parameters as in Section~\ref{res:numerics}. }
\end{table}
For illustration purposes, we take the medium parameters as $L=5$ fm and $T=0.4$ GeV and only show the band with coupling variation $g=1.8-2.2$. We do not show the fit uncertainities here, although we have checked that these are not larger than the error bands that we choose to show in Figure\ref{fig:nuratiofit}. From the fit values, we again note that the correlator with $\nu=0.01$ has the strongest angular dependence when compared to other $\nu$ values considered in the plot. 

\section{Full JEWEL simulation}
\label{apn:fulljewel}
In this section, we show the results of full JEWEL simulations for the ratios of $\nu$-correlators with that of the two-point correlator. For the plot shown in this section, we consider the default medium model encoded in JEWEL that incorporates a Bjorken-like expansion of the medium and consider the impact of recoil on the PE$\nu$C distributions. For medium distributions in the presence of recoil, we perform the constituent subtraction method for background subtraction as detailed in Ref.~\cite{KunnawalkamElayavalli:2017hxo}. 
\begin{figure}[h]
\centering 
\includegraphics[width=0.41\textwidth]{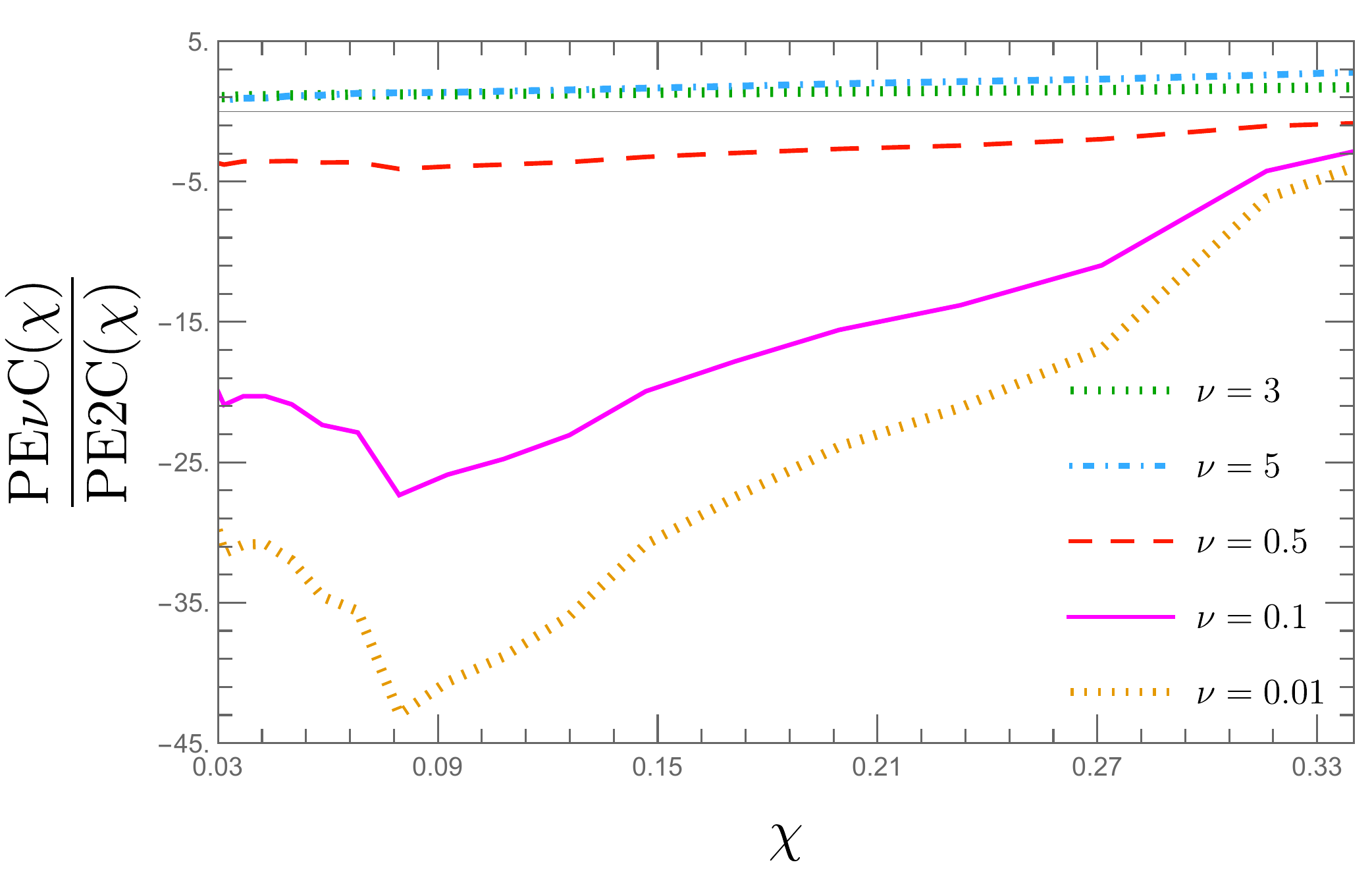}	
\caption{JEWEL simulation results incorporating medium recoil as well as hadronization effects.} 
\label{fig:nujewelfull}
\end{figure}

The full simulation results are shown in Figure~\ref{fig:nujewelfull}.
These results highlight that the features observed in our simple analytic setup for $\nu<1$ hold even in the light of a more realistic scenario. 

\section{Validation against asymptotic limits}
\label{apn:plotexp}

\begin{figure}[h]
\centering 
\includegraphics[width=0.43\textwidth]{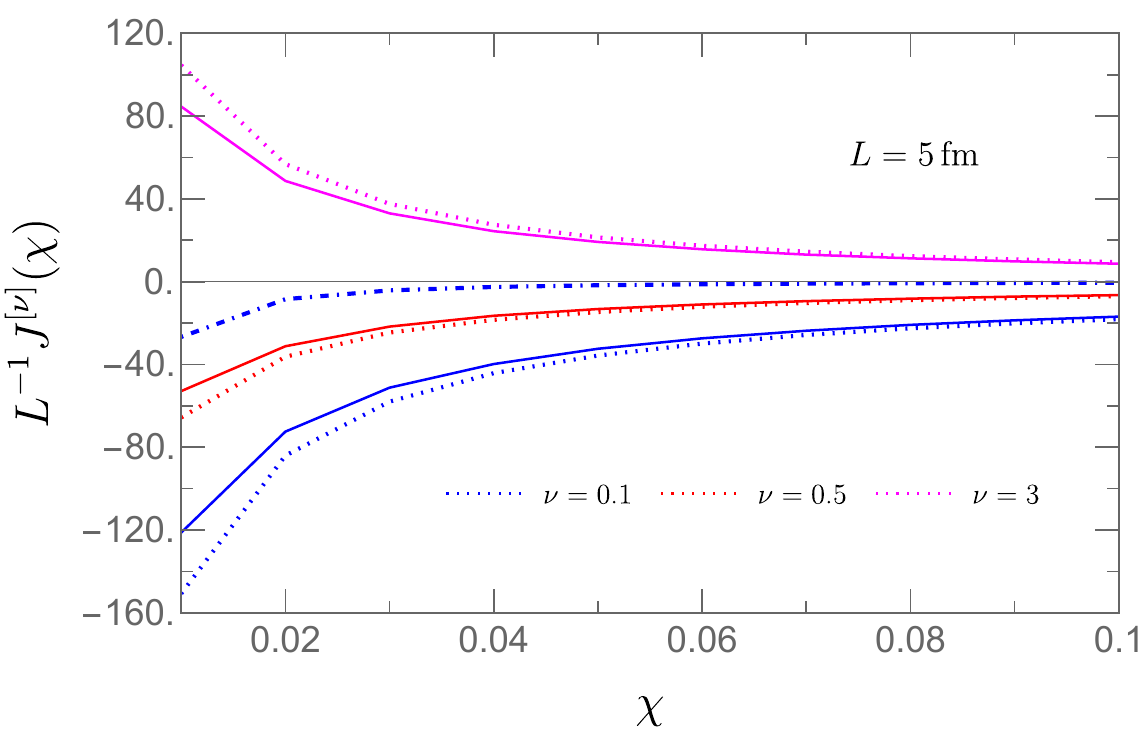}	
\caption{Comparison of the jet function over the full phase-space shown in Section~\ref{sec:results} against the asymptotic limits discussed in Section~\ref{sec:medium-jet-func}, as a function of angle $\chi$ and for various values of $\nu$. The medium parameters are $T=0.4$ GeV, $L=5$ fm and $m_D=0.8$ GeV. The solid lines represents full numerical results, the dotted lines denote $L\to\infty$ limit and the dotdashed line shows the distribution in $\kappa^2L\ll z\omega$. } 
\label{fig:nuchiapp}
\end{figure}

In this section, we present the comparison of our full numerical results against the asymptotic behaviors derived within the well-motivated approximations discussed in Section~\ref{sec:medium-jet-func}.  The results are shown in Figure~\ref{fig:nuchiapp} with $L = 5$ fm for both solid (full) and dotdashed ($\tau_f \geq L$ limit) while the dotted curves represent the limit $L \to \infty$. For comparison, we take medium function $\mathcal{B}\sim cg^4T^3/\bfk^4$ with $c=1$. 

From figure, we note that while $L \to \infty$ describes the full result reasonably well at angular scales with $\chi \geq  0.05$, the contribution from the limit $\tau_f \geq L$ is sufficiently suppressed for the whole range of $\chi$ values, contributing mostly in the small angle region. This suppression arises because the phase space for which the expansion is valid is very restrictive, i.e., $\vert \bfk \vert \ll 1/(2 \sqrt{\chi} L) \approx 0.2 $ GeV for $\chi \sim 0.01$, $L = 5$ fm and $T = 0.4$ GeV. Nevertheless, we observe that the full result consistently lies well between the two asymptotic limits, thereby validating our numerical estimates against the analytic forms derived in Section~\ref{sec:medium-jet-func}.

\bibliographystyle{elsarticle-num}
\bibliography{nupoint}

\end{document}